\documentstyle[12pt]{article}
\begin{document}

\voffset-1in
\footheight12pt\footskip30pt
\baselineskip15pt
\textheight 45 \baselineskip

\newsavebox{\uuunit}
\sbox{\uuunit}
    {\setlength{\unitlength}{0.825em}
     \begin{picture}(0.6,0.7)
        \thinlines
        \put(0,0){\line(1,0){0.5}}
        \put(0.15,0){\line(0,1){0.7}}
        \put(0.35,0){\line(0,1){0.8}}
       \multiput(0.3,0.8)(-0.04,-0.02){12}{\rule{0.5pt}{0.5pt}}
     \end {picture}}
\newcommand {\unity}{\mathord{\!\usebox{\uuunit}}}
\newcommand{\half}{{\textstyle\frac{1}{2}}}
%
\newcommand{\dr}{\raise.3ex\hbox{$\stackrel{\leftarrow}{\delta }$}}
\newcommand{\dl}{\raise.3ex\hbox{$\stackrel{\rightarrow}{\delta}$}}
%
\newcommand{\cA}{{\cal A}}
\newcommand{\cB}{{\cal B}}
\newcommand{\cC}{{\cal C}}
\newcommand{\cD}{{\cal D}}
\newcommand{\cE}{{\cal E}}
\newcommand{\cF}{{\cal F}}
\newcommand{\cG}{{\cal G}}
\newcommand{\cH}{{\cal H}}
\newcommand{\cI}{{\cal I}}
\newcommand{\cJ}{{\cal J}}
\newcommand{\cK}{{\cal K}}
\newcommand{\cL}{{\cal L}}
\newcommand{\cM}{{\cal M}}
\newcommand{\cN}{{\cal N}}
\newcommand{\cO}{{\cal O}}
\newcommand{\cP}{{\cal P}}
\newcommand{\cQ}{{\cal Q}}
\newcommand{\cR}{{\cal R}}
\newcommand{\cS}{{\cal S}}
\newcommand{\cT}{{\cal T}}
\newcommand{\cU}{{\cal U}}
\newcommand{\cV}{{\cal V}}
\newcommand{\cW}{{\cal W}}
\newcommand{\cX}{{\cal X}}
\newcommand{\cY}{{\cal Y}}
\newcommand{\cZ}{{\cal Z}}
%
\newcommand{\bA}{{\bf A}}
\newcommand{\bB}{{\bf B}}
\newcommand{\bC}{{\bf C}}
\newcommand{\bD}{{\bf D}}
\newcommand{\bE}{{\bf E}}
\newcommand{\bF}{{\bf F}}
\newcommand{\bG}{{\bf G}}
\newcommand{\bH}{{\bf H}}
\newcommand{\bI}{{\bf I}}
\newcommand{\bJ}{{\bf J}}
\newcommand{\bK}{{\bf K}}
\newcommand{\bL}{{\bf L}}
\newcommand{\bM}{{\bf M}}
\newcommand{\bN}{{\bf N}}
\newcommand{\bO}{{\bf O}}
\newcommand{\bP}{{\bf P}}
\newcommand{\bQ}{{\bf Q}}
\newcommand{\bR}{{\bf R}}
\newcommand{\bS}{{\bf S}}
\newcommand{\bT}{{\bf T}}
\newcommand{\bU}{{\bf U}}
\newcommand{\bV}{{\bf V}}
\newcommand{\bW}{{\bf W}}
\newcommand{\bX}{{\bf X}}
\newcommand{\bY}{{\bf Y}}
\newcommand{\bZ}{{\bf Z}}
\newcommand{\beq}{\begin{equation}}
\newcommand{\eeq}{\end{equation}}
\newcommand{\gras}[1]{\epsilon_{#1}}
\newcommand{\gh}[1]{\mbox{gh} \left( #1 \right)}
\newcommand{\sdet}{\mbox{sdet}}
\newcommand{\str}{\mbox{str}}
\newcommand{\tr}{\mbox{tr}}
\newcommand{\ihbar}{\frac{i}{\hbar}}
%
\newcommand{\sqrg}{\sqrt{g}}
\newcommand{\sqrabsg}{\sqrt{\vert g \vert}}
\newcommand{\ddr}{\raise.3ex\hbox{$\stackrel{\leftarrow}{d}$}}
\newcommand{\ddl}{\raise.3ex\hbox{$\stackrel{\rightarrow}{d}$}}

\begin{titlepage}
\begin{flushright} KUL-TF-93/44\\
                   hepth@xxx/9310166\\
                   October 1993\\
\end{flushright}
\vskip 5.mm
\begin{center}
{\large\bf  Construction of topological field theories using BV
}\\ \vskip 15.mm
{\bf F. De Jonghe $^1$ and S. Vandoren $^2$
}\\ \vskip 1cm
Instituut voor Theoretische Fysica
        \\Katholieke Universiteit Leuven
        \\Celestijnenlaan 200D
        \\B--3001 Leuven, Belgium\\[0.3cm]
\end{center}
\vskip 5.mm
\begin{center}
{\bf Abstract}
\end{center}
\begin{quote}
\small
We discuss in detail the construction of topological field theories using
the Batalin--Vilkovisky (BV) quantisation scheme. By carefully examining
the dependence of the antibracket on an external metric, we show that
differentiating with respect to the metric and the BRST charge do not
commute in general. We introduce the energy momentum tensor in this scheme
and show that it is BRST invariant, both for the classical and quantum
BRST operators. It is antifield dependent,
guaranteeing gauge independence.
For topological field theories, this energy momentum has to be quantum BRST
exact. This leads to conditions at each order in $\hbar$.
As an example of this procedure, we consider topological
Yang--Mills theory. We show how the reducible set of symmetries
used in topological Yang--Mills can be
recovered by means of trivial systems and canonical transformations. Self
duality of the antighosts is properly treated by introducing an infinite
tower of auxiliary fields. Finally, it is shown that the full energy
momentum tensor is classically BRST exact in the antibracket sense.
\vspace{4mm}  \hrule width 3.cm
{\footnotesize
\noindent $^1$ Aspirant N.F.W.O. Belgium, E--mail :
Frank\%tf\%fys@cc3.kuleuven.ac.be  \\
\noindent $^2$ E--mail : Stefan\%tf\%fys@cc3.kuleuven.ac.be}
\normalsize
\end{quote}
\end{titlepage}

\section{Introduction}
Topological field theories (TFT)\cite{Witten,report} have attracted a lot
of attention
recently. A general definition is that a TFT is characterised by the fact
that
its partition function is independent of the metric, which is considered to
be external and thus not included in the set of dynamical fields of the
theory~:
\begin{equation}
\frac{\delta Z}{\delta g^{\alpha \beta }}=\frac{\delta }{\delta g^{\alpha
\beta  }}\int {\cal D}\phi e^{\frac{i}{\hbar}S(\phi )}=0\ .
\end{equation}
If the action is BRST invariant, then this condition is satisfied
owing to the Ward identities, provided the
energy momentum tensor $\frac{2}{\sqrt |g|}\frac{\delta S}{\delta
g^{\alpha \beta }}$ is BRST exact.
Soon after their discovery by Witten,
topological field theories were shown \cite{construct,constructbv} to
be generally of the form
\begin{equation}
S=S_0+\delta _QV\ ,    \label{usual}
\end{equation}
where $S_0$ is either zero or a topological invariant (i.e. independent of
the metric) and BRST invariant. $\delta _QV$ is then the gauge fixing
term that corresponds to the shift symmetry of $S_0$. Using the formal
arguments of \cite{measure} based on Fujikawa variables to prove the metric
independence of the measure, one then has that
\begin{equation}
\frac{\delta Z}{\delta g^{\alpha \beta }}=\frac{i}{\hbar}\int {\cal D}\phi
\frac{\delta
}{\delta
g^{\alpha \beta  }} [\delta _QV]e^{\frac{i}{\hbar}S(\phi )}\ .
\end{equation}
Usually, differentiating with respect to the metric and taking the BRST
variation are freely commuted, which then leads to the desired result.
We will show in the next section, that the assumed commutation is not
allowed in general.

In order to investigate several steps of this process
in more detail, we will use the Batalin--Vilkovisky (BV)
\cite{BV,PWT,Bible} quantisation scheme. Although the BV scheme was used in
\cite{constructbv}
to gauge fix TFT \footnote{Recently \cite{shogo}, it was shown that the
$\Delta$ operator of BV appears in the Hamiltonian treatment of
topological sigma-models. This is not surprising, the BV scheme and its
generalisation to BRST--anti-BRST invariant quantisation of any theory have
recently
been derived starting from extending the usual BRST symmetries with
precisely shift symmetries \cite{ColFT}.}, the full
power of this scheme has apparently
not been exploited. In that scheme, one has to construct a quantum
extended action $W$ satisfying a quantum master equation
\beq
      (W,W) - 2i \hbar \Delta W = 0.     \label{QME}
\eeq
Formally, that is forgetting about the required regularisation
\cite{PWT}, one can then define a (quantum) BRST operator by
$\sigma X = (X,W) - i\hbar \Delta X$ which is nilpotent if $W$ satisfies
(\ref{QME}). Below we will define an energy-momentum tensor with the
property $\sigma T^q_{\alpha \beta} = 0$, by carefully specifying the
metric dependence of the antibracket and the $\Delta$-operator.
Hence, this $T^q_{\alpha \beta}$ is quantum BRST invariant.
For the theory to be topological, its energy momentum tensor
has to satisfy
\beq
             T^q_{\alpha \beta} = \sigma X_{\alpha \beta },
\eeq
which makes $T^q_{\alpha \beta}$ cohomologically equivalent
to zero. As both tensors appearing in this equation have an expansion in
$\hbar$,
this is a tower of equations, one for every order in $\hbar$. At the
classical level, we are looking for an $X_{\alpha \beta}^0$ such that
$T_{\alpha \beta} = (X^0_{\alpha \beta}, S)$, where $S$ is the classical
extended action of the BV scheme. We will show that non trivial
conditions appear for higher orders in $\hbar$. Even when no
quantum counterterms are
needed to maintain the Ward identity, the order $\hbar$
equation is non trivial.

In the next section, we will start by studying in detail the metric
dependence of the two typical operations in BV, i.e. the antibracket and
$\Delta$-operator. This immediately makes clear that taking the BRST
variation of $X$, i.e. calculating $(X,S)$ does not commute in general with
differentiating with respect to the metric. Furthermore, we define the
energy-momentum tensor mentioned above.
These are general results, not restricted to TFT. With this in mind,
we reconsider the construction of Topological Yang Mills theory by gauge
fixing. We show how working with the reducible set of symmetries (YM +
shift symmetry) can
be avoided in the BV scheme, as the YM transformation rules can always be
incorporated by introducing a trivial system and performing canonical
transformations. Also, we show how the selfduality of the auxiliary fields
can be treated carefully by introducing an infinite tower of auxiliary
gauge fixings. This does not spoil the topological nature of the theory.
Finally, we calculate the classical energy-momentum tensor following the
specified
prescription. Exploiting that canonical transformations leave both the
classical and quantum cohomologies invariant, we easily find an $X_{\alpha
\beta}$ such that $T_{\alpha \beta} = (X_{\alpha \beta},S)$. However,
this $X_{\alpha \beta}$ is not related to $V$ of (\ref{usual}).
All in all, this paper again shows the usefulness of
the antibracket formalism and especially its
canonical transformations.

\section{The energy-momentum tensor in BV}

We construct formally the energy-momentum tensor that
satisfies the condition $\sigma X = (X,W)-i\hbar \Delta X = 0$
or, classically only, $(X,S)=0$. In a first subsection, we will derive
expressions for the derivation of the antibracket and $\Delta$-operator
with respect to the metric. We
then define an energy-momentum tensor that satisfies the classical or
quantum condition of an observable. Finally, we show that this definition
is canonically invariant, which means that it is independent of the
gauge choice.

\subsection{Metric dependence of the antibracket and $\Delta$}
We
have to be precise on the occurences of the metric in all our expressions.
First of all, we have to specify a consistent set of conventions. All
integrations are with the volume element $dx\sqrabsg $. The functional
derivative is then defined as
\beq
   \frac{\delta \phi^A}{\delta \phi^B} = \frac{1}{\sqrabsg_B}
\frac{d\phi^A}{d\phi^B} = \frac{1}{\sqrabsg_B} \delta_{AB}\ ,
\label{FD}
\eeq
and the same for the antifields. The notation is that $A$ and $B$
contain both the discrete
and space-time indices, such that $\delta_{AB}$ contains both space-time
$\delta$-functions (without $\sqrabsg$) and Kronecker deltas ($1$ or
zero) for the discrete indices . $g$ is
$\det g_{\alpha \beta}$, and its subscript $B$ denotes that we evaluate it
in the space-time index contained in $B$.  Using this, the antibracket and
box operator are defined by\footnote{We then have that $(\phi ,\phi
^*)=\frac{1}{\sqrabsg}$. In this convention the extended action takes the
form $S=\int dx\sqrabsg
[{\cal L}_0+\phi ^*_i\delta \phi ^i+\phi ^*\phi ^*...]$. Demanding that
the total lagrangian is a scalar
amounts to taking the antifield of a scalar to be a
scalar, the antifield of a covariant vector to be a contravariant vector,
etc. One could also use the following set of conventions. We integrate
with the volume element $dx$ without metric, and define the functional
derivative (\ref{FD}) without $\sqrabsg$. Also the antibracket is defined
having no metric and so we have that $(\phi ,\phi ^*)'=1$. With this
bracket the extended action takes the form $S'=\int dx[\sqrabsg{\cal L}_0+
\phi ^*_i\delta \phi ^i+\frac{1}{\sqrabsg}\phi ^*\phi ^*...]$. The
relation between the two sets of conventions is a transformation
that scales the antifields with the metric, i.e. $\phi ^*\rightarrow
\sqrabsg
\phi ^*$, which makes them densities. In these variables, general
covariance
is not explicit and requires a good book--keeping of the $\sqrabsg$ 's in
the extended action and in other computations. Therefore, we will not use
this convention.}
\begin{eqnarray}
 ( A, B ) & = & \sum_i \int dx \sqrabsg_X \left( \frac{\dr A}{\delta
\phi^X} \frac{\dl B}{\delta \phi^*_X} - \frac{\dr A}{\delta \phi^*_X}
\frac{\dl B}{\delta \phi^X} \right) \nonumber \\
  \Delta A & = & \sum_i \int dx \sqrabsg_X (-1)^{\gras{X}+1}
\frac{\dr}{\delta \phi ^X}\frac{\dr}{\delta \phi^*_X} A\ ,
\end{eqnarray}
where $\gras{X}$ is the grassmann parity of the field $\phi ^X$.
For once, we made the summation that is hidden in the De Witt summation
more explicit. $X$ contains the discrete indices $i$ and the space-time
index $x$. These definitions guarantee that the antibracket of two
functionals is again a functional.
Using the notation introduced above, we have that
\begin{eqnarray}
 ( A, B ) & = & \sum_i \int dx \frac{1}{\sqrabsg_X} \left( \frac{\ddr
A}{d \phi^X} \frac{\ddl B}{d \phi^*_X} - \frac{\ddr A}{d \phi^*_X}
\frac{\ddl B}{d \phi^X} \right) \nonumber \\
  \Delta A & = & \sum_i \int dx \frac{1}{\sqrabsg_X} (-1)^{\gras{X}+1}
\frac{\ddr}{d \phi ^X}\frac{\ddr}{d \phi^*_X} A\ .
\end{eqnarray}
It is now simple to differentiate with respect to the metric. We use the
following rule~:
\begin{equation}
\frac{\delta g^{\alpha \beta }(x)}{\delta g^{\rho \gamma }(y)}=\frac{1}{2}
(\delta ^\alpha _\rho \delta ^\beta _\gamma +\delta ^\alpha _\gamma \delta
^\beta _\rho )\delta (x-y)\ ,
\end{equation}
where the $\delta $--function does not contain any metric, i.e.
$\int dx \delta (x-y)f(x)=f(y)$. This we do in order to agree with the
familiar recipe to calculate the energy-momentum tensor. Then we find that
\beq \label{metricbracket}
\frac{\delta (A,B)}{\delta g^{\alpha \beta}(y)}=\left( \frac{\delta
A}{\delta g^{\alpha \beta}(y)} , B \right) + \left( A , \frac{\delta
B}{\delta g^{\alpha \beta}(y)} \right)
+\frac{1}{2}g_{\alpha \beta }(y)\sqrabsg (y)[A,B](y)\ ,
\eeq
and
\beq
\frac{\delta \Delta A}{\delta g^{\alpha \beta}(y)}=\Delta \frac{\delta
A}{\delta g^{\alpha \beta}} +\frac{1}{2}g_{\alpha \beta }(y)\sqrabsg
(y)[\Delta A](y)\ ,
\label{metricdelta}
\eeq
with the notation that $(A,B)=\int dx\sqrabsg [A,B]$ and $\Delta A=\int
dx\sqrabsg [\Delta A]$. Notice that in $[A,B]$ and $[\Delta A]$ a
summation over the discrete indices is understood.
Before applying this to define the energy momentum tensor
in the BV scheme, consider the following properties, which we
will frequently use below. For any two operators $A$ and $B$, we have that
\beq
   \label{lemma1}
   \sum_i \phi^*_X \frac{\dl}{\delta \phi ^*_X}
   (A,B)=( \sum_i \phi^*_X \frac{\dl A}{\delta \phi ^*_X}, B)
   +( A , \sum_i \phi^*_X \frac{\dl B}{\delta \phi ^*_X})
   -[A,B](x)\ ,
\eeq
and
\beq
   \label{lemma2}
   \sum_i \phi^*_X \frac{\dl \Delta  A}{\delta \phi ^*_X} = \Delta \left(
   \sum_i \phi^*_X \frac{\dl A}{\delta \phi ^*_X}  \right)
   -[\Delta A](x)\ .
\eeq
In both expressions, $X=(i,x)$ with discrete indices $i$ and continuous
indices $x$. There is no integration over $x$ understood, only a summation
over $i$, which is explicitised.

Let us define the differential operator
\beq
     D_{\alpha \beta} = \frac{2}{\sqrabsg} \frac{\delta}{\delta g^{\alpha
\beta}} + g_{\alpha \beta} \sum_i \phi^*_X \frac{\dl}{\delta \phi^*_X}\ .
\eeq
Then it follows from (\ref{metricbracket}) and (\ref{lemma1}) that this
operator satisfies
\beq
    D_{\alpha\beta} (A,B) = (D_{\alpha\beta} A, B) + (A , D_{\alpha\beta}
B)\ .
\eeq
Owing to (\ref{metricdelta}) and (\ref{lemma2}), $D_{\alpha\beta}$ is seen
to commute with the $\Delta$-operator:
\beq
      D_{\alpha\beta} \Delta  A = \Delta D_{\alpha\beta} A\ .
\eeq
In particular, this shows that the BRST charge is metric dependent.
As
the BRST operator is simply the antibracket, i.e. $Q\,A(\phi)=(A,S)$,
$D_{\alpha \beta }$ does not commute with $Q$. This is not a consequence of
our conventions.

\subsection{Definition of the energy-momentum tensor}

Let us now apply all these results to define an expression which can be
interpreted as being the energy momentum tensor and that is invariant under
the BRST transformations in the antibracket sense. Define
\beq
      \theta_{\alpha \beta} = \frac{2}{\sqrabsg} \frac{\delta
S}{\delta g^{\alpha \beta}}\ .
\eeq
By differentiating the classical master equation $(S,S)=0$ with
respect to the metric $g^{\alpha \beta}(y)$, and by multiplying with
$2/\sqrabsg$, we find from (\ref{metricbracket}) that
\beq
   0 = 2 ( \theta_{\alpha \beta}(y) , S ) + 2 g_{\alpha \beta}(y)
\sum_i  \frac{\dr S}{\delta \phi^X} \frac{\dl S}{\delta
\phi^*_X}\ .
\eeq
In the second term, $X=(y,i)$ and there is only a summation over $i$.
Hence, we see that $\theta_{\alpha \beta}$ is not BRST invariant in the
antibracket sense, contrary to what one could naively expect.

However, if we define the energy-momentum tensor by
\beq
    T_{\alpha \beta} =  D_{\alpha\beta}  S,
\eeq
then it follows immediately that
\beq
    D_{\alpha\beta} (S,S) = 0 \Leftrightarrow ( T_{\alpha \beta} , S ) =0.
\eeq
It is then clear that $T_{\alpha \beta}$ is a BRST invariant
energy-momentum tensor.\footnote{
Notice that this quantity is the energy momentum tensor that one
immediately obtains when using the variables mentioned in the
previous footnote, i.e.
after scaling the antifields. One can then check that $T_{\alpha \beta
}=\frac{2}{\sqrabsg}\frac{\delta S'}{\delta g^{\alpha
\beta }}$. In this sense the modification of $\theta _{\alpha \beta
}$ is an artifact of the used conventions.}.
Moreover, $\theta_{\alpha \beta}\vert_{\phi^*=0} =
T_{\alpha \beta}\vert_{\phi^*=0}$.
Whether this
is a trivial element of the cohomology, i.e. equivalent to zero, can of
course not be derived on general grounds. By adding to this expression for
$T_{\alpha \beta}$ terms of the form $(X_{\alpha \beta},S)$, one can
obtain
cohomologically equivalent expressions. For example, by subtracting the
term $(\frac{1}{2} g_{\alpha \beta} \sum_i \phi^*_X \phi^X, S)$, the
terms that have to be added to $\theta_{\alpha \beta}$ to obtain $T_{\alpha
\beta}$ take a form that is symmetric in fields and antifields.

We can generalise this result and define an energy-momentum tensor that
is quantum BRST invariant. Consider
the quantum extended action $W$ that satisfies the quantum master equation
$ (W,W) - 2i\hbar \Delta W =0$. Define the quantum analogue of
$\theta_{\alpha \beta}$, i.e.
\beq
    \theta^q_{\alpha \beta} = \frac{2}{\sqrabsg}  \frac{\delta W}{\delta
g^{\alpha \beta}}\ .
\eeq
Again, one easily sees that this is not a quantum BRST invariant quantity.
Define however
\beq
     T^q_{\alpha \beta} =D_{\alpha\beta} W,
\eeq
then it follows by letting $D_{\alpha\beta}$ act on the quantum master
equation that
\beq
  \sigma T^q_{\alpha \beta} = (T^q_{\alpha \beta} , W ) - i\hbar \Delta
T^q_{\alpha \beta} = 0\ .
\eeq
Remember that $\sigma^2 = 0 $ when the quantum extended action satisfies
the quantum master equation and that hence $\sigma$ is to be considered the
quantum BRST operator.

\subsection{Canonical Invariance}

We will now show that our definition of the energy momentum tensor is
invariant under (infinitesimal) canonical transformations, up to a term
that is BRST exact.
Infinitesimal canonical transformations \cite{BVCT,PWT,Bible} are generated
by $ F(\phi,\phi^{*'}) = \phi^A \phi^{*'}_A +f(\phi,\phi^{*'})$,
with $f$ small. The transformation rules then become
\begin{eqnarray}
      \phi^{A'} & = &
      \phi^A + \frac{\delta f(\phi ,\phi^*)}{\delta \phi^*_A}
\nonumber \\
      \phi^{*'}_A & = &
      \phi^*_A - \frac{\delta f(\phi,\phi^*)}{\delta \phi^A}.
\end{eqnarray}
The expression in the primed coordinates for any function(al) given in the
unprimed coordinates can be obtained by direct substitution of the
transformation rules. Owing to the infinitesimal nature of the
transformation, we can expand in a Taylor series to linear order and we
find
\beq
   X'(\phi ',\phi^{*'})
       =  X (\phi^{A'} , \phi^{*'}_A ) - (X,f)\ .
\eeq
Especially, the
classical action and the energy-momentum tensor transform as follows:
\begin{eqnarray}
      S^{'} & = & S - (S,f) \nonumber \\
      T^{'}_{\alpha \beta} & = & T_{\alpha \beta} - (T_{\alpha \beta}, f)
      .
\end{eqnarray}
Here, $T_{\alpha \beta}$ is the energy-momentum tensor that is obtained
following the recipe given above starting from the extended action $S$.
Analogously, we can apply the recipe to the transformed action $S^{'}$,
which leads to an energy-momentum tensor $\tilde T_{\alpha \beta}$. Using
(\ref{metricbracket}) and (\ref{lemma1}), it is easy to show that
\begin{eqnarray}
  \tilde T_{\alpha \beta} &=&  D_{\alpha\beta} S' \nonumber \\
  & = & T^{'}_{\alpha \beta} - (S ,
\frac{2}{\sqrabsg} \frac{\delta f}{\delta g^{\alpha \beta}} + g_{\alpha
\beta} \sum_i \phi^*_X \frac{\dl f}{\delta \phi ^*_X}) \nonumber \\
  & = & T^{'}_{\alpha \beta} + (D_{\alpha\beta} f , S'),
\end{eqnarray}
as for infinitesimal transformations terms of order $f^2$ can be neglected.
We will use below that if $T_{\alpha \beta} = (X_{\alpha \beta}, S)$, then
$\tilde T_{\alpha \beta} = (\tilde X_{\alpha \beta}, S')$ as canonical
transformations do not change the antibracket cohomology. For
infinitesimal transformations we have that
\beq
\label{nice}
\tilde X_{\alpha \beta} = X_{\alpha \beta} - (X_{\alpha \beta},f)
+ \frac{2}{\sqrabsg} \frac{\delta f}{\delta g^{\alpha \beta}} + g_{\alpha
\beta} \sum_i \phi^*_X \frac{\dl f}{\delta \phi ^*_X}\ .
\eeq
We will use this result below.
We will finally verify that also $T^q_{\alpha \beta}$ is canonically
invariant. Under an infinitesimal canonical transformation, we have the
following transformation properties \cite{Bible}:
\begin{eqnarray}
       \tilde  W & = & W + \sigma  f = W' -i\hbar \Delta f
\nonumber \\
T^{q'}_{\alpha \beta} & = & T^q_{\alpha \beta} - (T^q_{\alpha
\beta},f),
\end{eqnarray}
with the same definition of $f$ as above. Notice that the solution of the
quantum master equation in the transformed set of coordinates is not $W'$
but $\tilde W$. The extra term is the logarithm of the Jacobian associated
with the transformation of the fields. Let us denote $\tilde T^q_{\alpha
\beta}$ the energy-momentum tensor that we obtain by applying the recipe to
the transformed action $\tilde W$. We then easily see that
\begin{eqnarray}
    \tilde T^q_{\alpha \beta} & = &  D_{\alpha\beta} \tilde W
 \nonumber \\
 & = & T^{q'}_{\alpha \beta} + \sigma \left[ \frac{2}{\sqrabsg}
\frac{\delta
f}{\delta g^{\alpha \beta}} + g_{\alpha \beta} \sum_i \phi^*_X \frac{\dl
f}{\delta \phi^*_X} \right].
\end{eqnarray}
Here too, rewriting the last term using $\sigma'$, the quantum BRST
operator in the transformed basis, only involves $f^2$ corrections.

\section{Topological Field Theories in BV}
After carefully introducing the energy momentum tensor, we define a
topological field theory by the condition
\begin{equation}
      T^q_{\alpha \beta} = \sigma X_{\alpha \beta}\ .
      \label{TME}
\end{equation}
Now we will prove that (\ref{TME}) implies
metric independence of the path integral. First we repeat that this condition
is gauge independent (canonical invariant), as was
explained in the previous section. For the path integral this means that
we may choose any gauge. This is done by doing a canonical
transformation such that in the new variables the new action is indeed
gauge fixed. After this transformation one may drop
the antifields~: \begin{equation}
Z=\int {\cal D}\phi e^{W(\phi, \phi ^*=0)}\,\, , \label{ME}
\end{equation}
where $W(\phi ,\phi ^*)$ is the quantum action satisfying the quantum
master equation.
For this path integral to be metric independent, we know that the following
condition has to be satisfied:
\begin{equation}
\int {\cal D}\phi\, \theta^q_{\alpha \beta }|_{\phi ^*=0}e^{W(\phi ,\phi
^*=0)}=0\ , \label{TE}
\end{equation}
where we used the notation of the previous section.
We also assumed that one can construct a metric independent measure,
which seems to
be possible in all known cases \cite{measure}. The above condition says
that the expectation value of the (quantum) energy momentum is zero.
One way the condition (\ref{TE}) may be satisfied, is by using
the Ward identity which in the BV scheme takes the form:
\begin{eqnarray}
0 & = &  \int \cD \phi \, \sigma X_A (\phi ,\phi^*)
   e^{\ihbar W(\phi ,\phi^*)} \nonumber \\
&=& \int \cD \phi  \, \left[ (X_A,W) - i \hbar \Delta X_A
\right] e^{\ihbar W(\phi ,\phi^*)}\ .
\end{eqnarray}
By the subscript $A$ we indicate that the Ward identity is valid,
whatever the indices of $X$ may be. Hence, we can prove that our theory is
topological, if we can find a quantity $X_{\alpha \beta}$ such that:
\begin{equation}
\theta^q_{\alpha \beta }|_{\phi ^*=0}=\sigma X_{\alpha \beta }|_{\phi
^*=0}
=(X_{\alpha \beta },W)|_{\phi ^*=0}-i\hbar \Delta X_{\alpha \beta }|_{\phi
^*=0}\ .
\end{equation}
As $T^q_{\alpha \beta }|_{\phi ^*=0}=\theta ^q_{\alpha \beta }|_{\phi
^*=0}$, then (\ref{TME}) certainly implies that the theory is
topological.

In general, both $W$ and $X_{\alpha \beta}$ have an expansion in terms of
$\hbar $~:
\begin{eqnarray}
W&=&S+\hbar M_1+\hbar^2M_2+\ldots \ .\nonumber \\
X_{\alpha \beta}&=&X^0_{\alpha \beta} + \hbar X^1_{\alpha \beta} +
\ldots \ .
\end{eqnarray}
Thus we see that (\ref{TME}) leads to a tower of equations, one for each
order in $\hbar$. The first two orders are
\begin{equation}
      T_{\alpha \beta} = (X^0_{\alpha \beta}, S)\ ,  \label{ctme}
\end{equation}
at the $\hbar^0$ level and
\begin{equation}
\frac{2}{\sqrabsg} \frac{\delta M_1}{\delta g^{\alpha \beta }}
+ g_{\alpha \beta} \sum_i \phi ^*_X \frac{\dl M_1}{\delta \phi^*_X}
=(X^0_{\alpha \beta },M_1)+(X^1_{\alpha \beta },S)-i\Delta
X^0_{\alpha \beta }\ ,\label{OLTME}
\end{equation}
at the one loop level.
Once $M_1$ is known from the one loop master equation \cite{PWT,Bible},
one has to solve (\ref{OLTME}) for $X^1$. This is an important
equation that must be satisfied at the one loop level.
To solve these equations one needs a regularisation
scheme, such that one can calculate (divergent) expressions like $\Delta S$
and $\Delta X^0_{\alpha \beta}$.  We hope to come back
to this issue in extenso somewhere else. If no (local) solution can be
found for $X_{\alpha \beta}$ then the proof of the topological nature
of the theory, based on the Ward identity, breaks down\footnote{In
\cite{report}, and references therein, it was shown that the one loop
renormalisation procedure
does not break the topological nature of the theory. However, the finite
counterterm $M_1$, to cancel eventual anomalies, and the $X^1$ term have
not been discussed.}. Notice that
even when no counterterm
$M_1$ is needed, one still has to solve the one loop equation if $\Delta
X^0_{\alpha \beta} \neq 0$.

Let us come back to the classical part of the discussion.
As is mentioned in the introduction, and as we will see in our example, the
gauge--fixed
action turns out to be BRST exact in the antibracket sense (up to a metric
independent term):
\begin{equation}
S=S_0+(X,S) ,     \label{SGF}
\end{equation}
where $S_0$ is a topological invariant. However, we want to stress
that this is not the fundamental equation to characterise TFT.
{}From this, it does not follow that (\ref{ctme}) is satisfied. Rather,
\beq
   T_{\alpha \beta} = ( \frac{2}{\sqrabsg} \frac{\delta X}{\delta g^{\alpha
\beta}} + g_{\alpha \beta} \sum_i \phi^*_X \frac{\dl X}{\delta \phi^*_X} ,
S ) + (X,T_{\alpha \beta})\ .
\eeq
In order for the theory to be topological, the second term should be BRST
trivial.

\section{Example~: Topological Yang--Mills}

We start from a manifold endowed with a metric
$g_{\alpha \beta }$ which may be of Euclidean or Minkowski signature. On
this manifold we define the Yang--Mills fields $A_\mu =A_\mu ^aT_a$.
The $T_a$ are the generators of a Lie
algebra. As the classical action\footnote{See
\cite{construct,constructbv}, and for pedagogical introductions
\cite{school}.}, we take the topological invariant known as the Pontryagin
index. So we have
\begin{equation}
S_0=\int _{\cal M}d^4x \sqrabsg F_{\mu \nu }\tilde {F}^{\mu \nu }\ .
\label{TYM}
\end{equation}
Here, the dual of an antisymmetric tensor $G_{\mu \nu }$ is defined by
\begin{equation}
\tilde {G}_{\mu \nu }=\frac{1}{2}[\epsilon ]_{\mu \nu \sigma \tau
}G_{\alpha \beta }g^{\alpha \sigma }g^{\beta \tau }\ .
\end{equation}
The Levi-Civita
tensor is defined by $[\epsilon ]_{\mu \nu \sigma \tau }=\sqrt{g}\epsilon
_{\mu \nu \sigma \tau }$, where $\epsilon _{\mu \nu \sigma \tau }$ is the
permutation symbol and $g=\det g_{\alpha \beta }$. Remark that it is
complex for a Minkowski metric. We normalise our reprensentation for the
algebra to be $Tr [T_a T_b] = \delta _{ab}$, and a trace over the
Yang-Mills indices is understood. The classical action
is invariant under continous deformations of the gauge fields that do not
change the winding number:
\begin{equation}
\delta A_\mu =\epsilon _\mu \ .  \label{SS}
\end{equation}

\subsection{Gauge fixing the action}

Following the approach of e.g. \cite{construct,constructbv,report} we will
now gauge fix this shift symmetry by
introducing ghosts $\psi _\mu $. Then we immediately
obtain the BV extended action
\begin{equation}
S=S_0+A^{*\mu}\psi _\mu \ . \label{S1}
\end{equation}
Remember that an overall $\sqrabsg$ is always understood in the volume
element of the space-time integration.
The usual approach is to include the
Yang--Mills gauge symmetry $\delta
A_\mu =D_\mu \epsilon $ as an extra symmetry, which then leads to a
reducible set of gauge
transformation as $D_\mu \epsilon $ is clearly a specific choice for
$\epsilon
_\mu $. We shall now argue that we can always introduce this
reducible symmetry
via a trivial system and canonical transformations.
This goes as
follows.
First, we enlarge the configuration space by introducing a fermionic ghost
field $c$. As it does not occur in
the extended action sofar, it can be shifted arbitrarily. For this
symmetry we introduce a ghost for ghost $\phi $.
The extended action then becomes
\begin{equation}
S=S_0+A^{*\mu}\psi _\mu+c^*\phi \ .
\end{equation}
Now we can do a canonical transformation, generated by the fermion
\begin{equation}
F=\unity-\psi '^{*\mu }D_{\mu }c\ ,
\end{equation}
which gives the transformation rules
\begin{eqnarray}
\psi _\mu &=&\psi '_\mu +D_\mu c\nonumber\\
c^*&=&c'^*+\partial _\mu \psi '^{*\mu }-\psi '^{*\mu }[A_\mu ,\cdot
]\nonumber\\
A^{*\mu }&=&A'^{*\mu }-\psi '^{*\mu }[\cdot ,c] \ .
\end{eqnarray}
The transformed action
is then (dropping the primes)~:
\begin{equation}
S=S_0+c^*\phi -\psi ^{*\mu }D_\mu (\phi -cc)+\psi ^{*\mu }(\psi _\mu
c+c\psi _\mu )+A^{*\mu }(\psi _\mu +D_\mu c)\ .
\end{equation}
Notice that the correct reducibility transformations have appeared in the
action, i.e. $A^\mu $ transforms under the shifts as well under Yang--Mills
. In order to make the connection to the usual (reducible)
procedure complete, we do yet another canonical transformation that
makes the familiar $c^*cc$ term of the Yang-Mills symmetry appear. This
transformation is generated by
\begin{equation}
G=\unity-\phi '^*cc \ .
\end{equation}
This gives $\phi '=\phi -cc$ and $c^*=c'^*-\phi '^*[c,\cdot ]$. After doing
these two canonical transformations, we have that
\begin{eqnarray}
S&=&S_0+A^{*\mu }(\psi _\mu +D_\mu c)+\psi ^{*\mu }(\psi _\mu c+c\psi _\mu
-D_\mu \phi )\nonumber\\&&+ c^*(\phi +cc)-\phi ^*[c,\phi ] \ .
\end{eqnarray}
Of course, this extra symmetry with ghost $\phi$, has to be gauge fixed
too. This is done in the literature by introducing a Lagrange
multiplier and antighost (sometimes called $\eta $ and $\bar \phi$).
As the BV scheme allows us
to enlarge the field content with trivial systems and perform canonical
transformations at any moment, we are free to choose to include them
or not. Therefore we drop them, keeping the gauge symmetries
irreducible.

Let us now gauge--fix the shift symmetry (\ref{SS}) in order to obtain the
topological
field theory that is related to the moduli space of self dual YM instantons
\cite{Witten}. We take the usual gauge fixing conditions
\begin{eqnarray}
\label{GF}
F^+_{\mu \nu }&=&0\nonumber\\
\partial _\mu A^\mu &=&0\ ,
\end{eqnarray}
where $G_{\mu \nu }^{\pm}=\frac{1}{2}(G_{\mu \nu }\pm \tilde {G}_{\mu \nu
})$. These projectors are orthogonal to eachother, so that we have
for general $X$ and $Y$ that $X^+_{\alpha \beta }Y^{-\alpha \beta }=0$. The
above gauge choice does not fix all the gauge freedom because there is no
unique solution of (\ref{GF})
for a given winding number. If that
would be the case
then the moduli space would consist out of one single point for every
winding number.
However, this gauge choice is admissible in the sense that the gauge fixed
action will have well defined propagators. Moreover, the degrees of freedom
that are left (the space of solutions of (\ref{GF})) form exactly the
moduli space of the instantons that
we want to explore. As in the usual BRST quantisation procedure, one has to
introduce auxiliary fields in order to construct a gauge
fermion. Obviously we should add
\begin{equation}
S_{nm}=S+\chi _{0\alpha \beta }^*\lambda _0^{\alpha \beta }+b^*\lambda \ ,
\label{S2}
\end{equation}
and consider the gauge fermion
\begin{equation}
\Psi _1=\chi _0^{\alpha \beta }(F^+_{\alpha
\beta }-x\lambda _{0\alpha \beta })+b(\partial _\mu A^\mu-y\lambda )\ ,
\label{CT1}
\end{equation}
where $x$ and $y$ are some arbitrary gauge
parameters. We introduced here an
antisymmetric field $\chi _0^{\alpha \beta }$. This field has
six components, which we use to impose three gauge conditions. So, we have
to constrain this field, e.g. by considering only self dual
fields $\chi _0^{\alpha \beta }= \chi _0^{+\alpha \beta }$. We can do
this as follows.
Our action after the canonical transformation (\ref{CT1}) has the gauge
symmetry $\chi _0^{\alpha \beta }\rightarrow \chi _0^{\alpha \beta
}+\epsilon
_0^{-\alpha \beta }$. So we fix this symmetry with the condition $\chi
_0^-=0$.
This can be done by adding an extra trivial system $(\chi _{1\alpha
\beta
},\lambda _{1\alpha \beta })$
and the canonical transformation with gauge fermion $F=\chi
_{1\alpha \beta }\chi _0^{-\alpha \beta }$. But then we have again
introduced too much fields,
and this leads to a new symmetry $\chi _{1\alpha \beta }\rightarrow
\chi _{1\alpha \beta }+\epsilon^+_{1\alpha \beta }$ which we have to
gauge fix. One easily sees that this procedure repeats itself ad infinitum.
We could, in principle, also solve this problem by only
introducing $\chi _0^{+\alpha \beta }$ as a field. Then we have
to integrate over the space of self dual fields.
To construct the measure on this space, we have to solve the constraint
$\chi =\chi ^+$.
Since this in general can
be complicated (as e.g. in the topological $\sigma
$--model) we will keep the $\chi _{\alpha \beta }$ as the
fundamental
fields. The path integral is with the measure ${\cal D}\chi _0^{\alpha
\beta }$ and we do not split this into the measures in the spaces of
self and anti--self dual fields. The price we have to pay is an infinite
tower of auxiliary fields. These we denote by
$(\chi _n^{\alpha \beta },\lambda _n^{\alpha \beta
})$\footnote{One remark has to be made here concerning the place of the
indices. We choose the indices of $\chi _n$ and $\lambda _n$ to be upper
resp. lower indices when $n$ is even resp. odd. Their antifields have
the opposite property, as usual.}
with statistics $\epsilon (\lambda _n)=n,\epsilon (\chi _n)=n+1$ (modulo
$2$) and ghost
numbers $gh(\lambda _n)$ equal to zero for $n$ even and one for $n$ odd.
Similarly, $gh(\chi _n)$ equals $-1$ for $n$ even and zero for $n$ odd. We
then add to the action (\ref{S2}) the term $\sum _{n=1}^{\infty}\chi
^*_{n,\alpha
\beta }\lambda _n^{\alpha \beta }$ and take as gauge fixing fermion
\begin{equation}
\Psi _2=\sum _{n=1}^{\infty}\chi _n^{\alpha \beta }\chi _{n-1,\alpha \beta
}^{(-)^{n}}\ + \Psi _1, \label{IJK}
\end{equation}
where $G_{\alpha \beta }^{(-)^n}$ denotes the self dual part of
$G_{\alpha
\beta }$ if $n$ is even and the anti self dual part if $n$ is odd.
After doing the gauge fixing we end up with
the following non--minimal solution of the classical master equation
\footnote{Note that from $(\chi ,\chi ^*)= \frac{1}{\sqrabsg}$, it follows
that
$(\chi ^{\pm},\chi ^{*\pm})=\frac{1}{\sqrabsg} P^{\pm}$ and $(\chi ^+,\chi
^{*-})=0$, where
$P^{\pm}$ are the projectors onto the (anti)-self dual sectors.}~:
\begin{eqnarray}
S_{nm}&=&S_0+A^{*\mu }\psi _\mu +(\partial _\mu A^\mu + b^*)\lambda
+( F^+_{\alpha \beta } + \chi^-_{1\alpha \beta } + \chi^*_{0\alpha \beta })
\lambda _0^{\alpha \beta }  \nonumber\\&&
- y\lambda ^2  -x\lambda _0^{\alpha \beta }
\lambda _{0\alpha \beta }+\chi _0^{+\alpha \beta }D_{[\alpha }\psi _{\beta
] }+b\partial _\mu \psi ^\mu \nonumber\\&&
+\sum _{n=1}^{\infty}(\chi ^*_{n\alpha \beta }+\chi
_{n+1,\alpha \beta }^{(-)^{(n+1)}}+\chi _{n-1,\alpha \beta
}^{(-)^{n}})\lambda _n^{\alpha \beta }\ .
\end{eqnarray}
We then obtain a gauge fixed action by putting all
antifields equal to zero. We now want to do the $\lambda _0$ and $\lambda$
integrals. Before doing that let us first make a comment on the measure.
Notice that we can construct a metric
independent measure, which can be shown by using the arguments of
\cite{measure}.
There the measure is written in terms of Fujikawa variables, giving a
Jacobian which is essentially $g^K$, for some number $K$. For our model,
one can easily show that $K=0$, yielding a metric
independent measure. Moreover, one can check that this independence is
still maintained after the
integrals over $\lambda _0$ and $\lambda$ have been done, if one keeps
track of
all determinants that appear when doing so. After these integrations, one
still has a solution of the classical master equation. This solution
is \begin{eqnarray}
S&=&S_0+\frac{1}{4y}(\partial _\mu A^\mu
+b^*)^2+\frac{1}{4x}(F^++\chi^-_1 +\chi
_0^*)^2\nonumber\\&&+b\partial _\mu \psi ^\mu +\chi _0^{+\alpha \beta
}D_{[\alpha }\psi _{\beta ]}+A_\mu ^*\psi ^\mu \nonumber \\
& &
+\sum _{n=1}^{\infty}(\chi ^*_{n\alpha \beta }+\chi
_{n+1,\alpha \beta }^{(-)^{(n+1)}}+\chi _{n-1,\alpha \beta
}^{(-)^{n}})\lambda _n^{\alpha \beta }\ .  \label{EA}
\end{eqnarray}
Notice that we now have terms quadratic in the antifields. This means that
the BRST operator defined by $Q\phi ^A=(\phi ^A,S)|_{\phi ^*=0}$ is only
nilpotent using field equations. Indeed, $Q^2b=\frac{1}{2x}\partial _\mu
\psi ^\mu \approx 0$, using the field equation of the field $b$.

We can write this extended action as $S=S_0 +(X,S)$, with $X$ given by
\beq
   X = \frac{1}{2} b (\partial_{\mu} A^{\mu} + b^*) - \psi_{\mu}^*
\psi^{\mu} + \frac{1}{2} \chi^{\mu \nu}_0 (F_{\mu \nu}^+ + \chi_{1 \mu
\nu}^- + \chi_{0\mu \nu}^* ) - \sum_{n=1}^{\infty} \lambda_n \lambda^*_n\ .
\eeq

\subsection{Energy momentum tensor}
We will now calculate the energy-momentum tensor, as defined in
section 2. As for notation, when a term is followed by $(\alpha
\leftrightarrow \beta)$, this means that this term, and only this one, has
to be copied but with the indices $\alpha$ and $\beta$ interchanged.
We find:
\begin{eqnarray}
T_{\alpha \beta }&=&
\frac{1}{2y}(\partial _\mu A^\mu +b^*)(\partial _\alpha
A_\beta +\partial _\beta A_\alpha )
\nonumber\\&&
-\frac{1}{4y}(\partial _\mu
A^\mu )^2g_{\alpha \beta }+\frac{1}{4y}(b^{*})^2g_{\alpha \beta
}-\frac{1}{8x}g_{\alpha \beta }F^2+\frac{1}{4x}g_{\alpha \beta }{\tilde
F}_{\mu \nu} \chi^{*\mu \nu} _0+\frac{1}{4x}g_{\alpha \beta }(\chi _0
^*)^2\nonumber\\&&+\frac{1}{4x}(F_{\alpha \nu }+\chi ^*_{0\alpha \nu
})(F_\beta {}^\nu +{\chi _0^*}_\beta{}^{\nu })+(\alpha \leftrightarrow
\beta )  \nonumber\\&&
+b(\partial _\alpha \psi _\beta +\partial _\beta \psi _\alpha
)-g_{\alpha \beta }b\partial _\mu \psi ^\mu
\nonumber\\&&-\frac{1}{2}g_{\alpha \beta }\chi^{\mu \nu}_0  D_{[\mu}
\psi_{\nu ]} -{\chi _0}^\rho
{}_\alpha {\widetilde {D_{[\rho }\psi _{\beta ]}}}-{\chi _0}^\rho {}_\beta
{\widetilde {D_{[\rho }\psi _{\alpha
]}}}\nonumber\\&&-\frac{1}{4x}g_{\alpha \beta }\chi _0^{*\mu \nu} {\tilde
\chi }_{1\mu \nu}  + \frac{1}{4x} (\chi_{1\alpha}{}^{\mu} +
\chi^*_{0\alpha}{}^{\mu} ) (\chi _{1\beta \mu} + \chi^*_{0 \beta \mu}) +
(\alpha \leftrightarrow \beta)-\frac{1}{8x}\chi _1^2\nonumber\\
&&- g_{\alpha \beta}
\sum _{n=1}^{\infty}[(\chi _{n+1,\mu \nu }^{(-)^{(n+1)}}
+\chi _{n-1,\mu \nu }^{(-)^{n}})\lambda _n^{\mu \nu }+\frac{1}{2}({\tilde
\chi }_{n+1,\mu \nu }-{\tilde \chi }_{n-1,\mu \nu })\lambda _n^{\mu \nu
}]\nonumber\\&&+\sum_{n=1}^{\infty}({\tilde \chi }_{n+1,\mu \alpha }-
{\tilde \chi }_{n-1,\mu \alpha }){\lambda _n}^\mu {}_\beta +
(\alpha \leftrightarrow \beta)\ .
\end{eqnarray}
We now determine $X_{\alpha \beta }^0$ such that
\begin{equation}
T_{\alpha \beta }=(X_{\alpha \beta }^0,S)\ , \label{CTME}
\end{equation}
which is the classical part of (\ref{TME}). Finding a solution of this
equation is a problem of antibracket cohomology. We could try to
construct the solution by an expansion in antifieldnumber, as explained in
\cite{STBV}, but this still turns out to be quite tedious. Instead, we will
take a different
strategy, using canonical transformations. We observe that (\ref{EA}) is
canonically equivalent to
\begin{equation}
S=S_0+A^{*\mu }\psi _\mu +\frac{1}{4y}(b^*)^2+\frac{1}{4x}(\chi _0^*)^2
+\sum_{n=1}^{\infty}\chi _{n,\alpha \beta }^*\lambda _n^{\alpha \beta }
\end{equation}
via a canonical transformation generated by a fermion $F=\unity + f$, with
\beq
         f = b\partial_{\mu}A^{\mu} + \chi_0^{\mu \nu} F^+_{\mu \nu} +
\sum_{n=1}^{\infty} \chi_n \chi_{n-1}^{(-1)^n}.
\eeq
Therefore, we can calculate the energy-momentum tensor in this set of
coordinates, verify that it is cohomologically trivial and transform the
result using (\ref{nice}).
For this simple action, we find
\begin{eqnarray}
X^{0}_{\alpha \beta }=\frac{1}{2}g_{\alpha \beta
}b^*b+\frac{1}{2}g_{\alpha
\beta }\chi _{0\mu \nu }^*\chi _0^{\mu \nu }+\chi ^*_{0\mu \alpha }{\chi
_0}^\mu {}_\beta\ .
\end{eqnarray}
Then it follows that in the new variables the solution is given by
\begin{eqnarray}
X^0_{\alpha \beta }&=&b(\partial _\alpha A_\beta +\partial _\beta A_\alpha
)-\frac{1}{2}g_{\alpha \beta }b\partial _\mu A^\mu +\frac{1}{2}g_{\alpha
\beta }b^*b\nonumber\\
&&+\chi _{0\mu \alpha }F^{-\mu }{}_\beta
+(\alpha \leftrightarrow
\beta )-\frac{1}{2}g_{\alpha \beta }\chi _0^{\mu \nu }F_{\mu \nu}^-
\nonumber\\
&&+\frac{1}{2}g_{\alpha \beta }\chi ^*_{0\mu \nu }\chi _0^{\mu \nu }+\chi
^*_{0\mu \alpha }{\chi _0}^\mu {}_\beta +(\alpha \leftrightarrow \beta
)\nonumber\\
&& + \frac{1}{2} g_{\alpha \beta} \chi_{1\mu \nu}^- \chi_0^{\mu \nu}
+ \chi^-_{1 \mu \alpha} \chi_0^{\mu} {}_{\beta} + (\alpha \leftrightarrow
\beta) \nonumber \\
&&-g_{\alpha \beta }\sum _{n=1}\chi _n( \chi _{n-1}^{(-)^n} + \frac{1}{2}
\tilde \chi_{n-1} ) + \sum_{n=1} \tilde \chi_{n \alpha \mu} \chi_{n-1
\beta }{}^{\mu} + (\alpha \leftrightarrow \beta ) \ .
\end{eqnarray}
One can indeed check that this expression satisfies (\ref{CTME}).
Notice that it contains $b^*b$ and $\chi^*_0 \chi_0$ terms.
Therefore, it is expected that the one loop equation (\ref{OLTME}) becomes
non-trivial.

\section{Conclusion}
In this paper we have shown how topological field theories
are described with the use of the BV language. An important role in these
theories is played by the energy momentum
tensor. The BV scheme has a natural energy-momentum tensor, obtained by
modifying the usual definition with some antifield dependent terms in order
to obtain
a gauge invariant quantity. Doing so, the energy momentum tensor is
automatically canonical invariant. Another important quantity is the BRST
charge. In the antifield formalism this is replaced by the
antibracket with the extended action. Then we proved that one can not
simply commute the BRST charge and the derivative with respect to
the metric.
We also sketched how to proceed for the full quantum theory and we showed
that the full (quantum)
energy momentum tensor, including all antifields, must be BRST exact
in order that the path integral is metric independent.
This leads to an equation
at each order of $\hbar$. It is not guaranteed that there exists a solution
of these equations. In any case, we expect these will give restrictions on the
theory.

As an example, we considered topological Yang--Mills theory. We
showed how to use the BV formalism to obtain the moduli space of instantons
by a proper gauge
fixing of the shift symmetry of the action (\ref{TYM}). In this framework,
one easily sees that
the Yang--Mills symmetry can be obtained by introducing an extra trivial
system and doing canonical transformations. This then makes the theory
reducible. At last, we gave the solution of the classical equation
mentioned above to prove the topological nature.

It is very promising to investigate the quantum theory for this model.
More generally, we think that topological field theories provide us with a
testing ground for a regularised treatment of the quantum BRST cohomology
of BV.

\section*{Acknowledgments}

We thank W. Troost, A. Van Proeyen and J.Par\'{\i}s for useful discussions.

\end{document}